\def\year{2021}\relax
\documentclass[letterpaper]{article} 
\usepackage{aaai20}  
\usepackage{times}  
\usepackage{helvet} 
\usepackage{courier}  
\usepackage[hyphens]{url}  
\usepackage{graphicx} 
\usepackage{graphicx}  
\usepackage{booktabs}  
\usepackage[compact]{titlesec}
\usepackage[authoryear,round,longnamesfirst]{natbib}

\title{Judging a book by its cover: Predicting the marginal impact of title on Reddit post popularity}

\author{
Evan Weissburg,
Arya Kumar,
Paramveer S. Dhillon\\
 University of Michigan\\ 
Ann Arbor, MI 48104\\
\{evancw$\vert$arkumar$\vert$dhillonp\}@umich.edu\\
{\bf (To appear at ICWSM 2022)}
}

\begin{document}


\maketitle

\begin{abstract}
Several factors influence the popularity of content on social media, including the {\it what}, {\it when}, and {\it who} of a post. Of these factors, the {\it what} and {\it when} of content are easiest to customize in order to maximize viewership and reach. Further, the title of a post (part of the {\it what}) is the easiest to tailor, compared to the post's body, which is often fixed. So, in this paper, we assess the impact of a post's title on its popularity while controlling for the time of posting (the {\it when}) by proposing an interpretable attention-based model. Our approach achieves state-of-the-art performance in predicting the popularity of posts on multiple online communities of various sizes, topics, and formats while still being parsimonious. Interpretation of our model's attention weights sheds light on the heterogeneous patterns in how the specific words in a post's title shape its popularity across different communities. Our results highlight the power of sentiment alignment, personal storytelling, and even personality politics in propelling content to virality.
\end{abstract}

\section{Introduction}
Online communities have an essential role in shaping public perception regarding critical societal issues. They serve as discussion forums and aid the dissemination of information. A crucial feature of most such communities is user feedback (both positive and negative) on posts, which determines content visibility and popularity. However, despite a deluge of social media posts, few reach a broad audience and become viral. Hence, it is crucial for social scientists to understand the determiners of content popularity, which could further shed light on the mechanisms by which online communities shape opinions. Social media platforms must also unpack the determiners of content popularity to improve the content they serve to users, as must content publishers that aim to produce compelling content.\footnote{Readers may recognize the task of content publishing when creating a picture caption for an Instagram post or crafting a humorous reply to a family member on Facebook.}

Although the problem is simple to formalize, modeling content popularity on social media is challenging because many factors determine popularity, including audience visibility, context, and time~\citep{PairwisePopPredict1}. These factors can be distilled down to three main elements of content, the {\it what} (what content was posted, e.g., the words, pictures, gifs, etc.), the {\it when} (when was the content posted, e.g., time-of-day, month, etc.), and the {\it who} (who posted the content, i.e., the identity of the author).

Of these three factors, the {\it what} and {\it when} can be easily customized to maximize viewership or reach, unlike the author ({\it who}), which is typically fixed. For instance, some content publishers post content on social media only at certain times, e.g., on Mondays or in the mornings, to garner as much widespread attention as possible. Similarly, it is common to tailor the title of a post to attract attention to the post's actual content body, which is often predetermined in the form of text, gif, image, or video to be shared. The post's title is also its most salient ``attention-grabbing'' attribute and is a crucial driver of a user's decision to click-through and consume the actual content. Further, titles are also a required component of posts across all social media websites and are, therefore, a generalizable property of such platforms. Hence, we focus on the title of a post in this paper.

\begin{figure}
  \centering
  \includegraphics[scale=0.3]{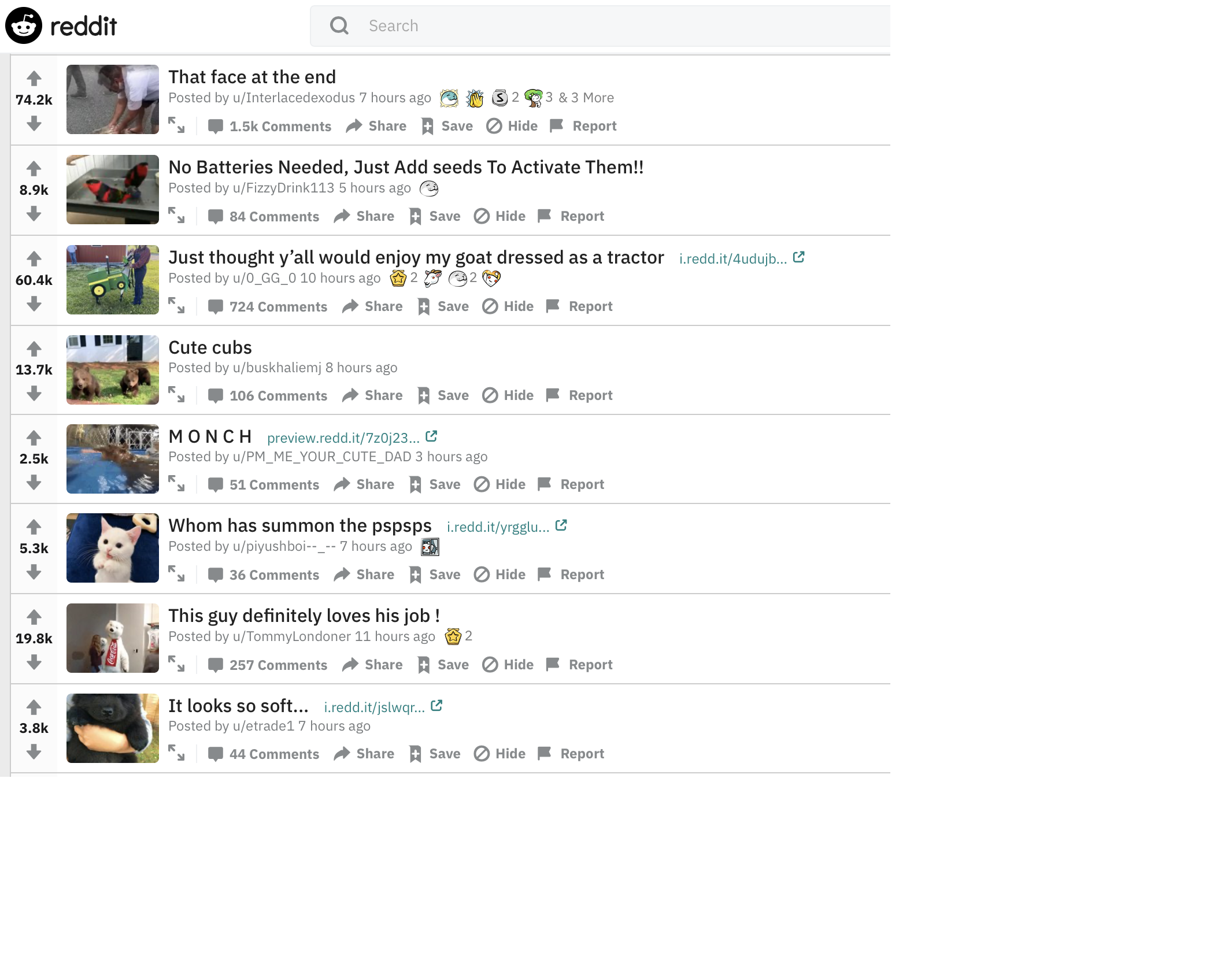}
  \vspace{-2cm}
  \caption{An example from subreddit /r/pics, where posts are ordered by user voting patterns. Older posts drop in position over time for freshness.}
  \label{fig:Model}
\end{figure}

This paper seeks to unbox the impact of a social media post's title on its popularity and further study the heterogeneous patterns across multiple online communities of various sizes, topics, and formats. We study the above research question on the \texttt{Reddit.com} social media community. We chose Reddit for our study since its size makes it highly representative of the overall internet. With many different topic-driven ``subreddit'' sub-communities, Reddit offers a rich landscape to study community dynamics in closer detail. These subreddit communities are centered around a single topic, so we can easily control for the general topic and audience by only comparing posts within the same subreddit. Figure~\ref{fig:Model} shows a snapshot of a subreddit.

In order to isolate the impact of the title of a post, we condition on the body of the post and assume it is fixed. We further consider only posts made within a 30-minute timeframe of each other to control for time ({\it  when}). The author's identity is less salient in determining content popularity on Reddit due to the lack of a significant ``follower'' mechanism. Hence, user identity effects are minimized, and controlling for the {\it who} is not warranted. Further, since we only compare posts from the same topic-driven subreddit community within a set time window, \textit{context} effects are also minimized.

Though we chose Reddit as the focus of our study, our model applies to any text or caption-oriented social media platforms, such as Twitter or Facebook. Our model could be trained at a platform level to recognize viral content or trained to learn the nuances of popular content on a smaller scale using specific community groups, as is the focus of our study.

Methodologically, we propose a simple novel neural network model that links the text of a post's title to its popularity. Our model extracts textual features by leveraging an attention mechanism~\citep{Attn}. Unlike deeper frameworks that use multiple attention layers totaling millions of parameters, we use only one layer to extract contextual clues from the post title~\citep{BERT}. This makes our model flexible, parsimonious, and interpretable.

Next, we interpret our results and model weights within each subreddit's context, which allows us to understand how important the title is to different communities on Reddit. Finally, we interpret the model's attention weights to determine which title words are most influential in propelling content to virality, illuminating the power of sentiment alignment, personal storytelling, and even personality politics. These interpretability analyses also reveal how a content publisher can adapt titles to a specific subreddit to boost its popularity potentially.

\section{Related Work}

Reddit, the sixth-largest website on the internet by unique traffic, is home to over a million subreddit communities. Each of these user-created themed communities is a microcosm of the broader social news ecosystem, aggregating thousands of user-generated posts sorted by realtime user voting feedback~\citep{AnatomyOfReddit}. Since Reddit provides such a user-centric slice of social media, it is a natural site to study online community interaction and content virality at scale~\citep{RMNReddit,ConvoModeling}. For example, many studies have attempted to predict {\it comment popularity} using linear regression, feed-forward neural networks, and recurrent neural networks (RNNs)~\citep{CommentPop2,CommentPop1,CommentPop3}. The most successful ones have used deep reinforcement learning, for instance, for predicting the popularity of future comment replies to existing threads~\citep{QPopPredict,ReinforcePopPredict}. All these studies reveal that, on Reddit, context is key to predicting popularity.

A different thread of research focuses squarely on predicting {\it post popularity}, a more difficult and input-rich task; rather than a simple sentence or paragraph, a user-generated post can include a link, image, gif, or video~\citep{ViralPsychology}. Additionally, many variables are involved in post popularity including title, content, creation time, subreddit of origin, and author~\citep{FactorsInPop2,ReinforcePopPredict,CatsAndCaptions,FactorsInPop3,TitlePopPredict,PairwisePopPredict1,FactorsInPop1}.\nocite{dhillon2012deterministic,dhillon2010inference,dhillon-etal-2012-metric} Disentangling the related effects of these variables to assess the weight of a single one is a difficult task, yet it often yields fruitful community insights.

Content popularity prediction is challenging in general because it is the result of a stochastic user-influenced positive feedback loop.~\citet{PredictionHard} explore this phenomenon, concluding that although the best content tends not to be rated poorly by validators and the least popular content tends not to be rated well, for most other posts, validation results were inconclusive. \citet{PostQuality} investigates this relationship between inherent post quality and post popularity further, introducing a viewership-corrected metric of inherent quality on dual datasets of Reddit and Hacker News. Contrary to~\citet{PredictionHard}, \citet{PostQuality} finds that his inherent quality metric has a strong correlation with popularity. However, \citet{PostQuality} omits a large quantity of data in his analysis, preferring to focus on high-ranking posts.

Similar content ``reposting'' is another phenomenon of interest, first studied in the context of post popularity prediction by~\citet{TitlePopPredict}. They developed a model based on title, image, and repost count to predict the relative popularity of reposted content generally on Reddit. Interestingly, they note that a repost following a previously popular high-visibility post is unlikely to be popular. 

Studies in other online communities have also found that title plays a critical role in content popularity. On a dataset of online news articles,~\citet{TitlesImportant} improve on popularity prediction baselines using a title-only model and propose that the content title is often ``the most prominent'' component of online content. Undoubtedly, content titles serve as useful hooks to draw initial attention.

Prior to our work, the most generalizable model for predicting post popularity on Reddit was a multimodal study on the role of captions versus images by~\citet{CatsAndCaptions}. They use existing machine learning architectures to contrast the importance of post titles and related images on subreddits \verb|r/aww|, \verb|r/pics|, and \verb|r/cats|. They show that treating popularity prediction as a pairwise task can be effectively used to control for time, so we also adopt this relative popularity prediction task in our model. Their results suggest that both text and image features are essential components in a popularity prediction model. Our results show conversely that popularity prediction is tractable without using image features and with a much smaller model. Therefore, our work differs in scope since we exclusively focus on the impact of a post's title in driving popularity and assume the post's content is fixed ahead of time, as is often the case when posting an image or link online. Further, while~\citet{CatsAndCaptions} analyze the role of caption versus image; our work focuses on presenting our state-of-the-art model for title-oriented popularity prediction, as well as developing a model-driven qualitative analysis of the community-level textual factors that propel content to virality.

\section{Problem Formulation}

In this section, we provide a brief overview of the Reddit platform and formalize the problem of post popularity prediction.

\subsection{Reddit Overview}

The community-driven discussion platform Reddit consists of multiple topic-focused subreddits. Within each subreddit community, users generate posts consisting of a title, content, and other associated metadata. Reddit allows registered users to upvote or downvote on posts, generating a per-post ``karma'' score, which determines a post's visibility~\citep{AnatomyOfReddit}. Reddit allows users to post a variety of multimodal content, though many subreddit communities place controls over post content; for example, \verb|r/AskReddit| allows no content (title only),  \verb|r/politics| allows only URL link-based submissions, \verb|r/pics| allows only images and \verb|r/depression| allows only text-based posts. We view a user-generated post $p \in \mathcal{P}$ as a collection of associated data: $p = ($score, title, content, subreddit, timestamp). While the user has no control over a post's score, they dictate its title, content, subreddit, and post creation time. 

\subsection{Problem Formalization}

As described earlier, we focus our popularity study on post titles rather than content. Titles feature most prominently in post previews where many users vote. Additionally, users have greater control over their post's title than their post's content \textemdash~ we focus on the common task of ``captioning'' where a news link or image content is already fixed~\citep{TwitterTitles}. This title-centric approach also allows us to contrast the role of post title across different subreddits by comparing our model's accuracy and attentional output between these communities. We are aware of no prior work that makes this type of deep cross-subreddit comparison.

Following the lead of several previous papers on predicting post popularity, we formulate the problem as a pairwise ranking task~\citep{ReinforcePopPredict,CatsAndCaptions,PairwisePopPredict1}. Given distinct posts $p_1, p_2 \in \mathcal{P}$, we aim to predict whether $p_1$ or $p_2$ has a higher associated popularity score using only the associated titles.

\section{Model}

\begin{figure}
  \centering
  \includegraphics[scale=0.33]{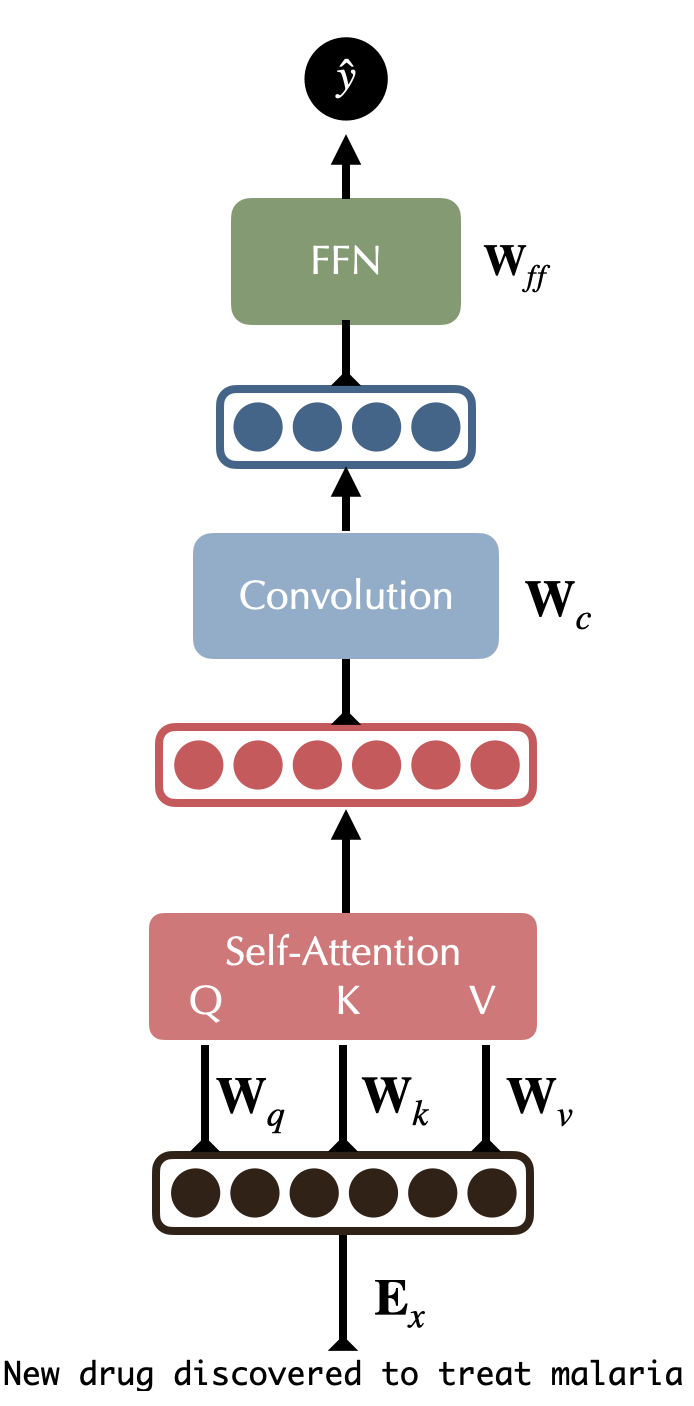}
  \caption{Model component illustration.}
  \label{fig:Model1}
\end{figure}

Our model is shown graphically in Figure~\ref{fig:Model1} and it consists of four layers: \textit {embedding}, \textit{attention}, \textit{convolution}, and \textit{feed-forward}. These layers are explained in more detail below.

\subsubsection{Embedding Layer:}

The embedding layer encodes the title information for a given post into fixed-length vectors. We use pretrained GloVe vectors for our embeddings, with embedding dimension $d_m = 300$~\citep{Glove}.

\subsubsection{Attention Mechanism:}

Though initially developed for sequence-to-sequence translation, the attention mechanism is now widely used for deep text-based natural language processing tasks, including studies with Reddit data~\citep{BERT,Attn,TomBERT,AttnUsed}. We simplify the fully stacked attention mechanism as used by~\citep{BERT,Attn} to a single self-attention module for our problem to make our model parsimonious. Through self-attention to word embeddings, this layer can add context to each word in the title, acting as a secondary embedding function serving input to the convolutional and dense layers described later.

The attention mechanism used in this paper is a version of scaled dot-product self-attention. The input consists of the fixed-length embeddings produced from the title, each of dimension $d_m$. These word embeddings are formulated into a matrix $M$, and the following attention product is computed.

\begin{equation}
  \mathrm{SelfAttention}(M) = \mathrm{softmax}\left(\frac{MM^T}{\sqrt{d_m}}\right)M
\end{equation}

\subsubsection{Convolutional Layer:}
Next, our model has a 1-dimensional convolutional layer, inspired by previous applications of convolution for text analysis~\citep{ConvEmb}. In this novel application, we convolve on attention outputs positionally with a fixed kernel size. Since we treat self-attention as a popularity contextualization function, inserting a convolutional layer between the attention output and the feed-forward module helps preserve positional information between word attention outputs.  For each window of length $k$, this mechanism processes input of dimension $(k, d_m)$ and produces a scalar. Overall, for an original input title with $n$ words and kernel size $k$, this mechanism produces vector output with length $d_n-k+1$. The convolutional layer also helps minimize per-token parameter counts before the final dense layer, thereby preventing overfitting.

\subsubsection{Feed-Forward Layer:}

The final module in our model is a feed-forward concatenation layer. This dense layer squashes the convolution layer's positional output into a scalar popularity prediction.

\subsubsection{Loss Function:}
Based on previous work in post popularity prediction, we use a pairwise ranking function to predict relative post popularity~\citep{Loss3,Loss1,CatsAndCaptions,Loss2} which lets us control for time effects since we compare posts that were posted within a 30-minute timeframe. To incentivize predictive confidence, we would like to maximize the pairwise difference in model outputs, so we implement a contrastive max-margin loss function. Given two posts $p_1, p_2$ such that $\mathrm{score}(p_1) > \mathrm{score}(p_2)$, with model outputs $x_1, x_2$, we compute,

\begin{equation}
  \mathrm{Loss}(x_1, x_2) = \mathrm{max}(0, x_2-x_1)
\end{equation}

For inference purposes, it is also possible to interpret the model's output $x$, for an unpaired post $p$.

\section{Experimental Setup}

This section describes the training and evaluation details of our model and the Reddit post selection/pairing algorithm for our pairwise-ranking loss function. Our model and baselines were trained on an Nvidia 2080Ti with 512 GB RAM. Our model processed approximately 100 pairs per second without batching and around 3000 pairs per second with a batch size of 64. Our source code is accessible online\footnote{https://github.com/evanweissburg/judging-a-book/}.

\subsection{Dataset Source}

Our Reddit data is sourced from \verb|pushshift.io|, an online dataset generated using Reddit's public API. This dataset was originally scraped in 2015, extended in 2016, and made public in 2020~\citep{Pushshift,Pushshift2,Pushshift1}.

We focus our attention on Reddit submissions from 2017. For each subreddit of interest, we filter posts according to the following criteria. We remove posts that received few (less than 2) upvotes and the posts that were ``stickied'' on the subreddit\textemdash a mechanism that allows subreddit moderators to artificially boost the visibility of posts circumventing the normal voting-based process. Table~\ref{tab:SubSize} shows the details of the sixteen subreddits that we used for our study.\footnote{For computational reasons, subreddit /r/The\_Donald was reduced in size by 1/6 by random sampling.}

\begin{table}
  \begin{center}
  \begin{tabular}{cc}
    \hline
        \hline
   Subreddit & Post Pairs \\
    \hline
    /r/Showerthoughts & 44,863 \\
    /r/AskReddit & 47,313 \\
    /r/news & 27,692 \\
    /r/worldnews & 38,576 \\
    /r/relationships & 21,962 \\
    /r/depression & 8,787 \\
    /r/aww & 103,604 \\
    /r/pics & 52,628 \\
    /r/politics & 142,218 \\
    /r/The\_Donald & 103,774\\
    /r/sports & 3,019 \\
    /r/soccer & 40,807 \\
    /r/science & 4,957 \\
    /r/NoStupidQuestions & 6,828 \\
    /r/funny & 92,699 \\
    /r/jokes & 29,971 \\
  \hline
  \hline
 \end{tabular}
 \caption{Dataset Size by Subreddit\label{tab:SubSize}}
 \end{center}
\end{table}

\subsection{Preliminary Analysis}

\begin{figure}
    \centering
    \includegraphics[scale=0.33]{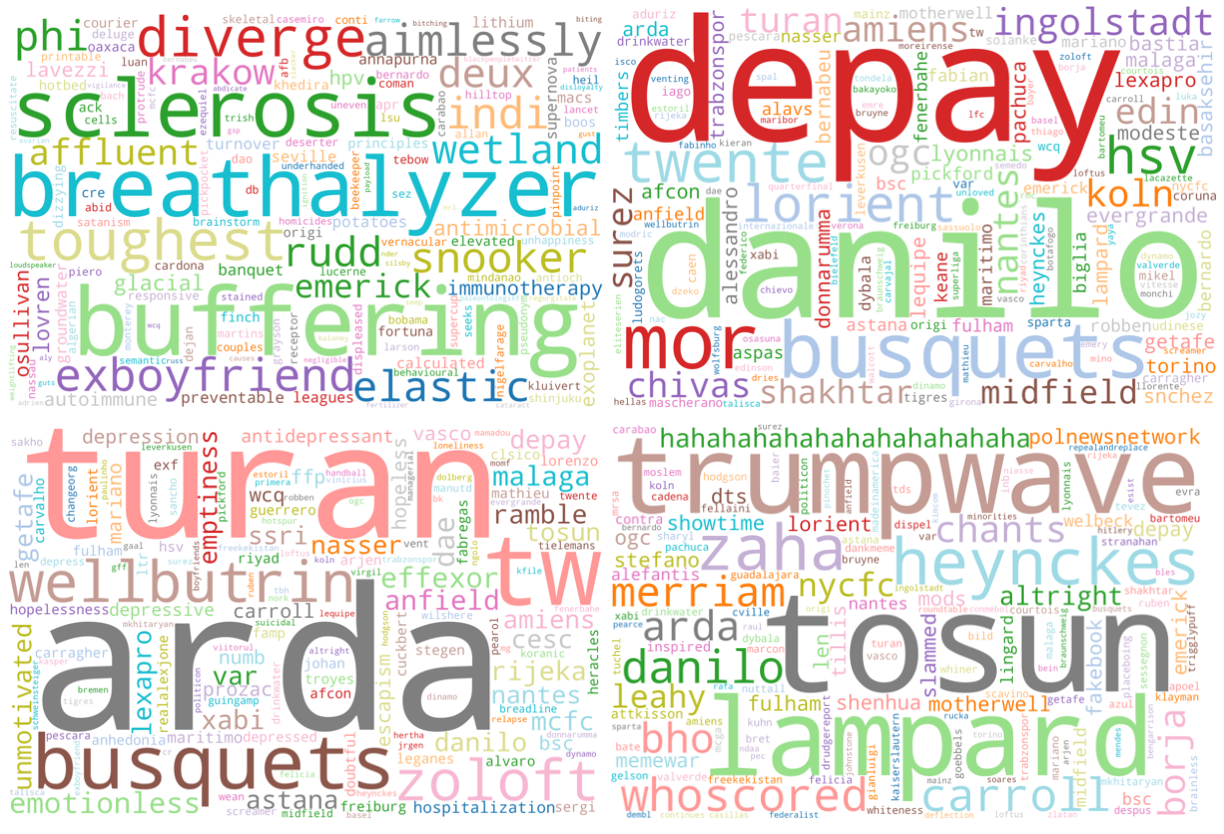}
    \caption{Word cloud visualization of (un)popular content on Reddit stratified by quartiles; Top Left = 75-100\% (most popular), Top Right = 50-75\%, Bottom Left = 25-50\%, Bottom Right = 0-25\%}
    \label{fig:WordClouds}
\end{figure}

To provide a preliminary view into the types of content that is popular on the sixteen subreddits that we study, we display word cloud visualizations, split into quartiles based on popularity. The score of any word $w$ is computed as the average scores of the posts $p \in \mathcal{P}$ in which it appears. To prevent larger subreddits from dominating the results, each post's score is normalized by the mean of the top 100 posts in the subreddit. Additionally, to avoid misspelled words in low scoring posts, we only show words in the graphic that appear at least 10 times in the corpus. Once each word has an associated score, we generate a word cloud for the 150 random words whose score falls into the corresponding quartile, as shown in Figure \ref{fig:WordClouds}.

At first glance, words that appeal to high-stakes events and emotions perform favorably, from medical terms (``autoimmune'', ``sclerosis'') to descriptions of relationships and trauma (``exboyfriend'', ``homicides'', ``unhappiness''). We see references to famous individuals, like Arsenal striker Pierre-Emerick Aubameyang, and to shared frustrations, like slowly-buffering content. In the bottom quartile, we see that certain controversial sentiment (``trumpwave'', ``fakebook'', ``whiteness'') appears to score poorly overall. Similarly, simple rules such as ``soccer players score well'' do not appear to be true \textemdash~ names of soccer players appear in every quartile of the results, emphasizing that constructing a viral post on Reddit requires nuance.

\subsection{Post Pairing}

Our post pairing algorithm controls for audience and time to minimize confounding in the dataset. First, since Reddit subreddits vary in viewership, we only pair posts from the same subreddit. To control for time, we greedily select posts ordered by time so that posts in a pair are from a similar time window. Since gaps can exist in a subreddit's posting history due to temporary closures or site unavailability, we also guarantee posts must be sourced within a 30-minute time frame. Following~\citet{CatsAndCaptions}, we minimize noise by pairing posts if their voting score difference is at least 20 and the more popular post's score is at least double the score of the less popular post.

\subsection{Baseline Models}

We implement four strong baselines against which we compare our model, as illustrated in Table~\ref{tab:ModelParams}. First, we implement a simple logistic regression using one-hot sentence encoding and a vocabulary of 20,000 words. With its low parameter count, this model provides a simple and interpretable baseline result. Next, we implement a two-layer multilayer perceptron model (MLP) that takes a GloVe sentence vector average as input. Our next baseline model is a more sophisticated deep learning method for textual data\textemdash a bidirectional LSTM model (BiLSTM) that takes GloVe word vectors as input.  Finally, as is standard practice these days, we provide a fine-tuned BERT baseline for the task~\citep{BERT}. We used BERT-Base from the Huggingface\footnote{https://huggingface.co/transformers/model\_doc/bert.html} library and added a 2-layer dense module to the output. To make a fair comparison, we freeze all but the final dense layers for training purposes. We also benchmarked against a light-weight BERT variant (DistilBERT), but the results were very similar to BERT-Base, so we only report the BERT-Base results. Although BiLSTM and BERT-Base generally outperform weaker baselines, their outputs are difficult to interpret due to their vast number of parameters. In comparison, our model has approximately 1/4 the number of parameters as BiLSTM.

\begin{table}
{\small
  \begin{center}
  \begin{tabular}{ccccc}
     \hline
        \hline
   1Hot Logistic & GloVe MLP & BiLSTM & BERT-Base & Ours \\
    \hline
    20k & 2M & 600k & 110M & 500k \\
   \hline
        \hline
\end{tabular}
 \caption{Parameter Count by Model\label{tab:ModelParams}}
 \end{center}
 }
\end{table}

Finally, we perform 5-fold cross-validation for all our experiments. We randomly shuffle the data (subreddit post pairs) and train our model on 4/5 of the data and test on the remaining 1/5. This also allows us to measure variability in our results and hence also assess statistical significance. 

\section{Results}
\begin{table*}[htbp]
\centering
  \begin{tabular}{ccccccc}
    \hline
        \hline
    Submission Type & Subreddit & 1Hot Logistic & GloVe MLP & BiLSTM & BERT-Base & Our model \\
    \hline
    Title-only & /r/Showerthoughts & 0.577 \small{(0.004)} & 0.583 \small{(0.008)} & 0.593 \small{(0.006)} & 0.596 \small{(0.004)} & \textbf{0.603 \small{(0.001)}} \\
    & /r/AskReddit & 0.607 \small{(0.004)} & 0.628 \small{(0.003)} & 0.632 \small{(0.003)} & 0.630 \small{(0.004)} & \textbf{0.644 \small{(0.004)}} \\
    \hline
    Link & /r/news & 0.595 \small{(0.006)} & 0.605 \small{(0.004)} & 0.613 \small{(0.005)} & 0.602 \small{(0.006)} & \textbf{0.637 \small{(0.005)}} \\
    & /r/worldnews & 0.633 \small{(0.008)} & 0.638 \small{(0.006)} & 0.650 \small{(0.001)} & 0.647 \small{(0.004)} & \textbf{0.663 \small{(0.002)}}  \\
    \hline
    Text & /r/relationships & 0.695 \small{(0.005)} & 0.732 \small{(0.004)} & 0.748 \small{(0.006)} & 0.747 \small{(0.003)} & \textbf{0.756 \small{(0.004)}} \\
    & /r/depression & 0.649 \small{(0.010)} & 0.706 \small{(0.010)} & 0.733 \small{(0.005)} & 0.704 \small{(0.006)} & \textbf{0.741 \small{(0.006)}} \\
     \hline
    Image & /r/aww & 0.597 \small{(0.003)} & 0.594 \small{(0.004)} & \textbf{0.609 \small{(0.003)}} &  0.596 \small{(0.003)} &\textbf{0.609 \small{(0.001)}} \\
    & /r/pics & 0.596 \small{(0.007)} & 0.605 \small{(0.004)} & \textbf{0.618 \small{(0.005)}} & 0.594 \small{(0.003)} & \textbf{0.618 \small{(0.003)}} \\
  \hline
        \hline
\end{tabular}
\caption{Model Results by Subreddit Submission Type. {\it Note:} 5-fold cross validation accuracy is reported with standard deviation in parenthesis; results significant at the $\alpha=0.05$ level in a pairwise t-test are bolded.\label{tab:SubSubmissionType}}
\end{table*}

\subsection{Accuracy}
Table~\ref{tab:SubSubmissionType} presents our results on eight popular subreddits divided into four categories by {\it subreddit submission type}. We note that our model performs best compared to baselines on subreddits where the content submission type is title-only and link, as expected. Overall, our model is competitive across all content types but is not statistically distinguishable in terms of its accuracy on subreddits with image content, matching BiLSTM in both cases. This suggests that the post's title is a less important determiner of its popularity when the content that it points to is image-based.

Next, we present another eight subreddits in Table~\ref{tab:SubTopic}, divided into four categories by topic. Here, we note that our model is effective on a wide range of subreddit discussion topics and vocabularies.

\begin{table*}[htbp]
\centering
  \begin{tabular}{ccccccc}
    \hline
        \hline
    Topic & Subreddit & 1Hot Logistic & GloVe MLP & BiLSTM & BERT-Base & Our model \\
    \hline
    Politics & /r/politics & 0.612 \small{(0.005)} & 0.631 \small{(0.002)} & 0.638 \small{(0.004)} & 0.616 \small{(0.003)} & \textbf{0.642 \small{(0.004)}} \\
    & /r/The\_Donald & 0.637 \small{(0.003)} & 0.631 \small{(0.004)} & 0.640 \small{(0.003)} & 0.634 \small{(0.001)} & \textbf{0.652 \small{(0.001)}} \\
    \hline
    Sports & /r/sports & 0.511 \small{(0.012)} & 0.589 \small{(0.012)} & 0.617 \small{(0.024)} & 0.577 \small{(0.013)} & \textbf{0.640 \small{(0.006)}} \\
    & /r/soccer & 0.635 \small{(0.003)} & 0.642 \small{(0.007)} & 0.651 \small{(0.010)} & 0.625 \small{(0.005)} & \textbf{0.661 \small{(0.003)}} \\
    \hline
    Science & /r/science & 0.620 \small{(0.009)} & 0.641 \small{(0.010)} & 0.674 \small{(0.011)} & 0.673 \small{(0.013)} & \textbf{0.695 \small{(0.015)}} \\
    & /r/NoStupidQuestions & 0.532 \small{(0.013)} & 0.594 \small{(0.014)} & 0.611 \small{(0.003)} & 0.599 \small{(0.009)} & \textbf{0.634 \small{(0.005)}} \\
     \hline
    Humor & /r/funny & 0.560 \small{(0.004)} & 0.565 \small{(0.004)} & 0.574 \small{(0.003)} & 0.566 \small{(0.005)} & \textbf{0.580 \small{(0.002)}} \\
    & /r/jokes & 0.586 \small{(0.004)} & \textbf{0.591 \small{(0.009)}} & \textbf{0.596 \small{(0.006)}} & 0.590 \small{(0.003)} & \textbf{0.600 \small{(0.004)}}\\
  \hline
        \hline
\end{tabular}
\caption{Model Results by Subreddit Topic. {\it Note:} 5-fold cross validation accuracy is reported with standard deviation in parenthesis; results significant at the $\alpha=0.05$ level in a pairwise t-test are bolded.\label{tab:SubTopic}}
\end{table*}

\subsection{Ablation Study}
Though our model is relatively simple, we performed an ablation study on the subreddit \verb|/r/travel| to determine the efficacy of the convolutional layer. It turns out that the predictive accuracy of our model drops significantly from 0.823 to 0.808 if we remove the convolutional layer, which shows its importance in terms of contributing to sparsity and minimizing overfitting on smaller subreddits.

\subsection{Interpreting Attention Weights}

Next, we interpret our model's output attention weights. Since this is not a prediction task, we train our model on the entire dataset. For an input title of length $k$, the corresponding model self-attention output weights have shape $(k, k)$. 

Researchers have used attention mechanisms to understand model behavior in tasks such as recommender systems, neural machine translation, and text labelling~\citep{AttnInterpret1,AttnInterpret2,AttnInterpret3,AttnNMT,AttnMedData,AttnSomething}. Theoretically, we frame our qualitative analysis as a view into the words and phrases that the downstream model is most interested in. We perform three types of qualitative studies interpreting relative attention weightings, as outlined below. 

\subsubsection{Top Subreddit Attention Weights}
Looking at the top few attention weights provides a simple but broad insight into popular content on a given subreddit. For each subreddit, the 15 words with highest attention weight are reported in Table~\ref{tab:Top15BySub}.

Though the performance of the 1Hot Logistic regression baseline is inferior to our model, it is similarly easy to interpret. Thus, Table~\ref{tab:LogisticTop} reports the 20 top (absolute) feature weights from the 1Hot Logistic model as well as the top-20 attention weights output by our model for the \verb|/r/politics| dataset. As can be seen, there is minimal overlap in the two sets of word distributions, which highlights the ability of our model to learn highly discriminative words to predict post popularity.

\subsubsection{Title Weights by Subreddit Model}

\begin{figure*}
  \centering
  \includegraphics[scale=0.4]{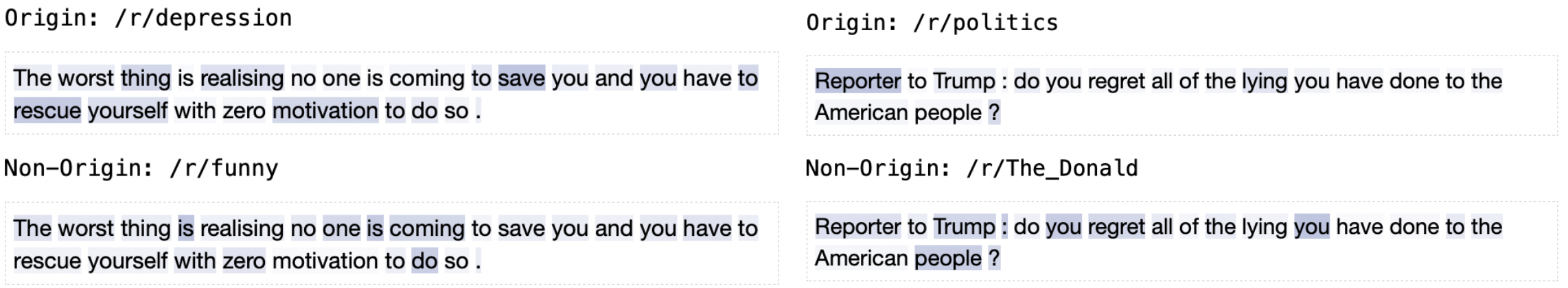}
  \caption{Comparing attention weights generated by two different subreddit models for the same title: ``Save'' and ``rescue'' contribute to popularity on /r/depression, but not on /r/funny ({\it left panel}). Similarly, references to ``Trump,'' ``regret,'' and use of the second person boost popularity more on /r/The\_Donald than on /r/politics ({\it right panel}).}
  \label{fig:CompareFunnyDepression}
\end{figure*}

To understand subreddit popularity more deeply, it is helpful to visualize the same title with models trained on different subreddits. In Figure~\ref{fig:CompareFunnyDepression} we choose a viral title from some {\it origin} subreddit and examine differences in word-level attention weighting using a non-origin subreddit model. By analyzing the same title using model variants trained on different subreddits, we gain insight into how our model interprets community-level variation in popular content. For example, we might expect our model to pay more attention to ``Trump'' in \verb|/r/The_Donald| than \verb|/r/politics|, and we observe this behavior in the right-panel of Figure~\ref{fig:CompareFunnyDepression}.

\begin{table}[h]
{\small
  \begin{tabular}{c|c|c|c}
    \hline         \hline
   /r/AskReddit & /r/relationships & /r/depression & /r/pics\\
    \hline 
    underwear & surgery & isolate & married \\
    9/11 & update & hug & weirdest \\
    yelp & injury & convince & upvotes \\
    woods & disability & dog & diet \\
    nsfw & son & killed & marry \\
    panties & option & cat & lbs \\
    bridges & stalking & cried & filmed \\
    sfw & therapy & kill & redditors \\
    flash & twins & pet & questions \\
    sleeper & property & picture & answers \\
    pg-13 & terrified & hardest & lb \\
    wwe & creepy & terrified & filming \\
    tortured & devastated & crush & diagnosed \\
    erection & autistic & coward & answer \\
    briefs & bathroom & knew & skyrim \\
  \hline   \hline
\end{tabular}
\caption{Top-15 words according to our model's attention weights for different subreddits\label{tab:Top15BySub}.}
}
\end{table}

\begin{table}[h]
{
  \begin{center}
  \begin{tabular}{cc|cc}
    \hline         \hline
   1Hot Logistic & Weight & Our model & Weight \\
    \hline 
    Korea & -0.22 & Las & 4.49 \\
    North & -0.21 & Venezuela & 3.99 \\
    poll & 0.14 & Vietnam & 3.90 \\
    report & 0.14 & Olbermann & 3.63 \\
    Trump & 0.13 & WikiLeaks & 3.60 \\
    Donald & 0.13 & transcript & 3.56 \\
    Fox & 0.13 & Warren & 3.50 \\
    neutrality & 0.13 & MSNBC & 3.48 \\
    dem & 0.12 & neutrality & 3.40 \\
    China & -0.12 & WSJ & 3.40 \\
    US & -0.12 & marijuana & 3.19 \\
    Russia & 0.11 & Fox & 3.11 \\
    Syria & -0.11 & look & 3.04 \\
    net & 0.11 & Assange & 3.03 \\
    immigration & -0.11 & Maddow & 3.01 \\
    court & -0.11 & impeached & 3.00 \\
    Supreme & -0.11 & EU & 3.00 \\
    travel & -0.11 & explains & 2.96 \\
    americans & 0.10 & tells & 2.92 \\
    Jeff & 0.10 & trial & 2.91 \\
  \hline   \hline
\end{tabular}
 \caption{Comparing the top-20 words based on their weights for the baseline 1Hot Logistic and our model on /r/politics\label{tab:LogisticTop}}
 \end{center}
}
\end{table}

\subsubsection{Subreddit Attention Directed Graphs}

A more powerful way to compare subreddit communities at scale using our model is by generating an attention directed graph, as in Figures~\ref{fig:ADGDepression}-\ref{fig:ADGJokes}. These graphs are generated from the trained subreddit model by collecting average attention weights for each input word-output word pair. To create each visualization, we systematically graph the strongest average input-output attention weights as directed edges, with stopwords removed for visual clarity. Therefore, a directed edge from word A to word B implies that our model believes A is an influential context for B with respect to the task of popularity prediction.

\section{Discussion}

As shown in Tables~\ref{tab:ModelParams},~\ref{tab:SubSubmissionType}, and~\ref{tab:SubTopic}, our approach consistently beats strong baselines by 1-3\% for the pairwise prediction task with a comparably small number of parameters. Hence, our model achieves both state-of-the-art performance and parsimony.

Previously top-performing baselines like BiLSTM and BERT-Base preclude model interpretability and are highly overparameterized. On the other hand, our approach combines high accuracy with weight interpretability akin to logistic regression. But unlike a logistic regression, our models' attention weights can be analyzed more deeply at the community level to uncover contextual trends in online popularity, as we explore below.

\begin{figure*}[!ht]
  \centering
  \includegraphics[scale=0.5]{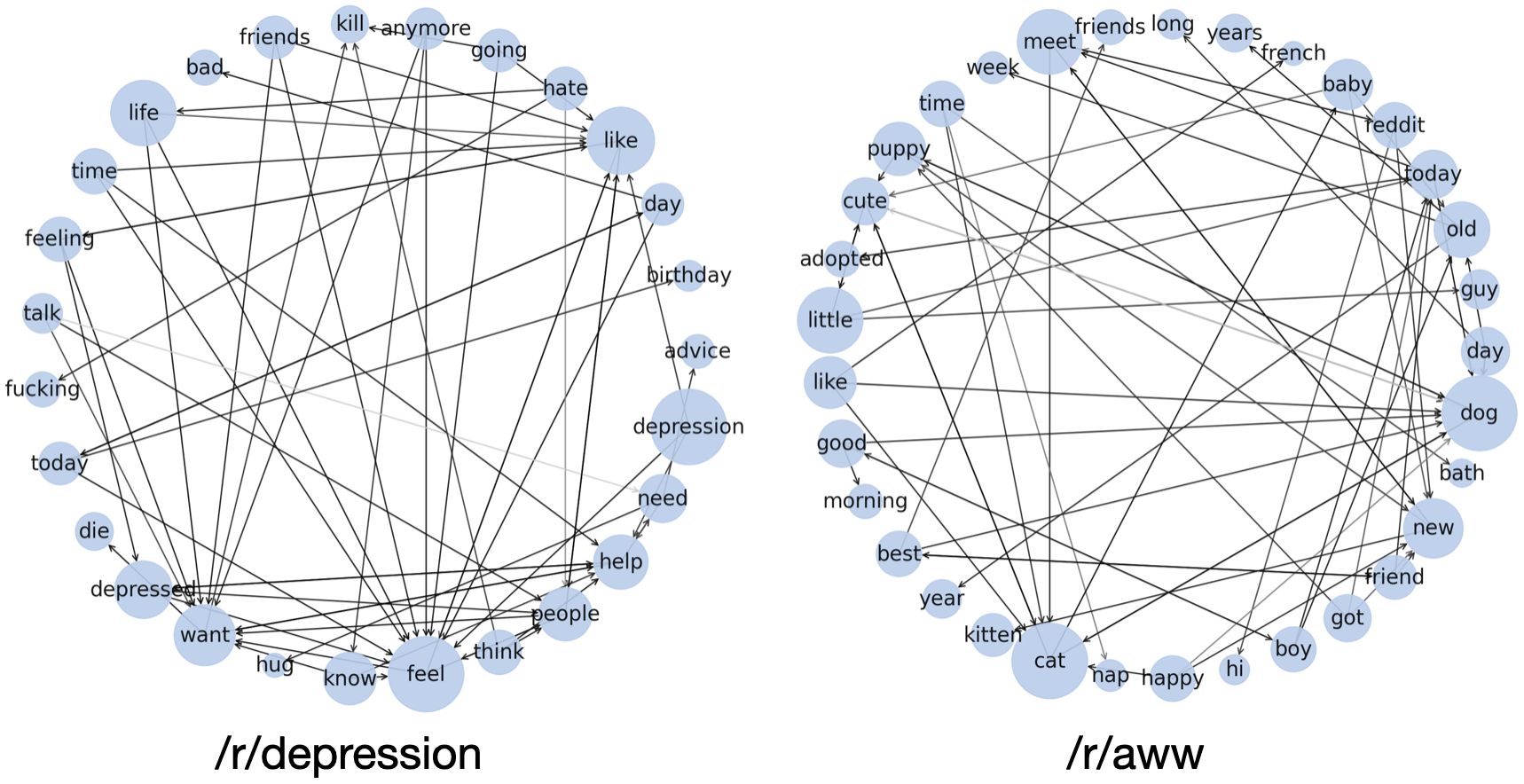}
  \caption{Subreddit graphs generated from attention weights showing which word nodes the popularity model links to others via directed edge; \textbf{(Left Figure)} Word Graph for /r/depression; \textbf{(Right Figure)} Word Graph for /r/aww.\label{fig:ADGDepression}}
\end{figure*}

It is worth emphasizing that our study's primary objective is not to produce a more accurate popularity prediction model for Reddit. If that were the desiderata, then we would augment our model with more complex features rather than just the title. Instead, our focus is to capture the differential impact of post title across various communities and uncover interesting popularity patterns, as we discuss next.

Table~\ref{tab:SubSubmissionType} shows, albeit a bit unsurprisingly, that our model demonstrates the most significant improvement over baselines on title-only and link-based subreddits. We expect it to perform best on title-only subreddits since, by definition, our model is title-oriented. These results also point to a secondary result: post title is also crucial on link-based subreddits. A plausible explanation for this is that Reddit provides an easy mechanism to vote on a link-based post after only viewing the post title. Therefore, many Redditors may fail to read externally-linked material before voting~\citep{ReadWhatShare}. Next, we look at performance of our model on subreddits which contain topical content. As can be seen in Table~\ref{tab:SubTopic}, we find that title is a particularly important factor in popularity for science and sports-related content. Interestingly, our approach does not demonstrate considerable improvements on humor-related subreddits.

In Tables~\ref{tab:Top15BySub} and \ref{tab:LogisticTop}, we list the words to which the model attributes the highest attention weight; theoretically, per subreddit, these words are the most effective captioning tools to improve the popularity of content. Table \ref{tab:LogisticTop} also provides a comparison between our model's top words and the baseline 1Hot Logistic regression, which yield slightly different results. On \verb|/r/politics|, we find only two words of twenty (``neutrality'' and ``fox'') are the same between both models, as shown in Table \ref{tab:LogisticTop}. While both models emphasize the importance of key names such as ``Donald Trump,'' ``Warren,'' or ``Maddow,'' our model places additional significance on personal opinions, such as ``explains'' and ``tells''.

Exploring differences in language between subreddits through these word weights yields community insight. Previous work in online content virality found that popular content is mainly propelled by its positive valence and physiological arousal. For example, posts that inspire awe, rage, or anxiety tend to be more viral than posts that create a deactivating feeling of sadness~\citep{DisgustingSharing,DrivesOnlineVirality}. Our qualitative findings align well with these hypotheses, though the relative roles of valence versus arousal differ between communities.

On \verb|/r/AskReddit|, a subreddit based on asking opinion-based questions of the broader community, it is evident from Table~\ref{tab:Top15BySub} that the most viral posts reference emotional extremes such as fear, violence, and sexual content. \verb|/r/relationships| indicates a similar trend: posts about creepy, stalking, or abusive relationships engender the most upvotes and community engagement. Based on these observations, we conclude that both of these communities tend to promote content that produces high arousal, whether through eroticism, anger, or a sense of injustice.

The subreddit \verb|/r/depression| tells a slightly different story. Although high-arousal negative words like ``isolate'' and ``killed'' are common, lower-arousal positive valence words that indicate support systems are also prevalent. From ``hug'' to ``cat,'' ``dog,'' or ``pet,'' this community clarifies that online post popularity allows for a full range of emotion. 

An unexpected trend comes from the picture-sharing subreddit \verb|/r/pics|. Instead of words indicating interest, beauty, or landscape, this community primarily popularizes pictures regarding personal health and weight loss. Although this subreddit may have begun as ``[a] place for pictures and photographs,'' per its current description, it appears to have evolved in focus through user-curation, a community migration phenomenon unique to Reddit. Rather than being drawn to exciting or beautiful images in isolation, users of \verb|/r/pics| prefer pictures that come with emotional positive personal narratives, like overcoming cancer, losing weight, or getting married \citep{StoriesMatter}.

\begin{figure*}
  \centering
  \includegraphics[scale=0.4]{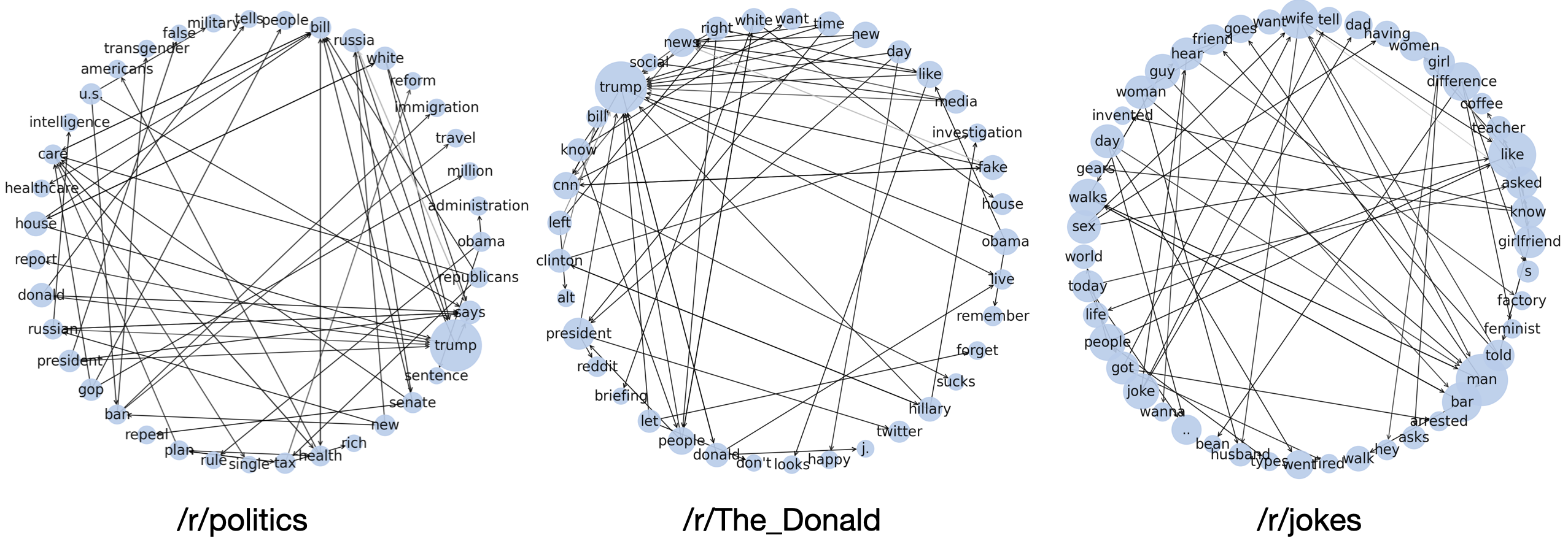}
  \caption{Subreddit graphs generated from attention weights showing which word nodes the popularity model links to others via directed edge; {\bf (Left Figure)} Word Graph for /r/politics; {\bf (Middle Figure)} Word Graph for /r/The\_Donald; {\bf (Right Figure)} Word Graph for /r/jokes}
  \label{fig:ADGPolitics}
  \label{fig:ADGJokes}
\end{figure*}


When the simple top-N word weights are insufficient insight, our model allows deeper analysis at the subreddit level using an attention-directed word graph. \verb|/r/depression| shows a range of emotions in popular posts, as indicated in Figure \ref{fig:ADGDepression}. One can visually note certain themes from the graph vocabulary and connection network: loneliness (``want talk [to] people,'' ``need friends/hug,'' ``today [is my] birthday'') and depression (``feel depressed,'' ``want kill/die'') are associated with the strongest attentional word connections, indicating popular subreddit sentiment. Compared to the word-level attention weights in Table \ref{tab:Top15BySub} which show a significant number of viral positive words, we find fewer positive valence word chains in this figure. Since our model identifies fewer common positive valence word chains, we infer that within \verb|/r/depression|, popular positive-valence posts tend to be more complex and varied in sentence structure than comparatively popular negative-valence posts, which appear to revolve around the same word chains and topics. This narrative is supported by Figure~\ref{fig:CompareFunnyDepression}, an example comparing attention outputs for a title originating on \verb|/r/depression|. We find that ``save you'' and ``rescue yourself'' are strong popularity boosters on \verb|/r/depression|, whereas the \verb|/r/funny| model variant's attention output is distributed evenly, indicating that these negative-valence word chains would fail to boost post popularity on \verb|/r/funny|.

The attentional graph for \verb|/r/aww| lies in sharp contrast with \verb|/r/depression|, indicating substantial differences in the format and relative complexity of subreddit title construction. The largest nodes reflect universally cute imagery (cats and dogs, puppies, kittens, and babies). With \verb|r/aww| in particular, strong word chains become clear (``Reddit, meet my cute new puppy,'' or ``Good morning from my happy kittens''), indicating the simplicity and repetitive nature of successful titles. Here in particular, the value of self-attention is clear; with these visualizations, we gain insight into particular phrases and pairings that are successful on a given subreddit. In \verb|/r/aww|, for example, cats and dogs are frequently used together, and ``new'' is usually used in conjunction with ``Reddit'' and ``meet''.

Similar patterns can be found for \verb|/r/jokes| in Figure \ref{fig:ADGJokes}. Just as with \verb|/r/aww/|, there are clear word chains (``A man walks into a bar'' or ``Wanna hear a joke''). However, the graph as a whole highlights an interesting issue. Many of the most common words in the titles are verbs, and these connect almost exclusively to subjects \textemdash~ the most salient features our model can extract here are the actions that the subject of the joke takes. However, often the most crucial element of a joke is a pun or clever use of word choice. These elements can be difficult to capture with the GloVe word embeddings that our model uses, since these vectors struggle to accurately represent multiple meanings of a given word at once.

Finally, we use our model outputs to provide insight into political discussion on both sides of the aisle. In our analysis of \verb|/r/politics| (a subreddit for US politics that typically leans liberal) and \verb|/r/The_Donald| (a subreddit for supporters of Donald Trump), we find significant thematic differences in attentional word associations. Though both subreddits attribute what ``Trump says'' to content popularity,  \verb|/r/politics| focuses on Trump's standpoints on political issues whereas \verb|/r/The_Donald| is more interested in Trump's opinions about figures and institutions. Note that the strongest connections surrounding Trump on \verb|/r/politics| are ``bill,'' ``immigration,'' and ``healthcare,'' compared to ``CNN,'' ``Clinton,'' ``Obama,'' and the ``media'' on \verb|/r/The_Donald|. Our model highlights that those who frequent the pro-Trump \verb|/r/The_Donald| promote content that centers around Trump's personality rather than his political agenda, compared to the more agnostic \verb|/r/politics|. In Figure \ref{fig:ComparePoliticsDonald}, we see a comparative attention visualization of an August 2020 news title exemplifying further differences. While \verb|/r/politics| seems to be interested in the ``reporter'' and potential ``lying,'' \verb|/r/The_Donald| is intrigued that ``Trump,'' ``regret[s],'' ``you,'' and ``people'' are mentioned, indicating that the model trained on \verb|/r/The_Donald| believes this populist sentiment will help propel this particular post to virality within the scope of its community.

\section{Conclusion}
In this paper, we unpacked the marginal impact of the title of a social media post in driving its popularity while controlling for the post body and the post's timing. We leveraged a self-attention based model to extract salient features from post titles. Our model showed strong performance on various subreddits while being parsimonious. Further, the model's attention weights permit an in-depth analysis of individual posts and community trends beyond simply identifying popular keywords. We compared online communities at scale through subreddit-level attention visualizations, highlighting differences in sentiment, word choice, and political ideology. Through our per-subreddit qualitative analysis, we find that viral content is extremely heterogenous on Reddit. This emphasizes the importance of considering the audience of a target community in order to tailor the text of a title; while an uplifting and positive narrative might be effective in generating popular content within one community, a caption arousing frustration might perform better in another. Our work provides an intuitive attention-based framework to study inter-community differences in detail.

Future work on online popularity prediction could consider the addition of new features to maximize performance, such as submission time, image or text content, or authorship in the context of an ensemble-type model. It might also be desirable to adapt and apply our model to other text-oriented platforms, such as Twitter or Facebook. For example, it could be enlightening to obtain a cross-platform view of content popularity to compare and contrast viral content at an internet scale on Reddit versus Instagram. Since our model is parsimonious and attention-based, new interpretation or visualization techniques could also be developed in future work to obtain novel insights into content virality.

\section{Ethics Considerations}

This study was conducted on public data, in accordance with the Reddit Terms of Service. However, users may not expect their data to be used and disclosed for research purposes. We acknowledge and mitigate this consideration by focusing our analysis on broad community-level phenomena, with the goal of understanding trends in online interaction. 

Another ethical consideration is use of this research for tailoring misinformation or propaganda. These issues make it more critical to analyze the factors involved in online content virality in order to understand and combat these dangers.

\bibliographystyle{aaai}
\bibliography{icwsm21_e_weissburg}

\end{document}